\documentclass[osajnl,twocolumn,showpacs,superscriptaddress,10pt]{revtex4-1} 
\usepackage{amsmath,amssymb,graphicx,bbm}
\usepackage{color}

\begin{document}

\title{Phase modulated multiphoton microscopy }

\author{Khadga Jung Karki}\email{Khadga.Karki@chemphys.lu.se}
\affiliation{Chemical Physics, Lund University, Getingev\"agen 60, 222 41, Lund, Sweden}
\author{Mohammed Abdellah}\affiliation{Chemical Physics, Lund University, Getingev\"agen 60, 222 41, Lund, Sweden}
\author{T\~onu Pullerits}\affiliation{Chemical Physics, Lund University, Getingev\"agen 60, 222 41, Lund, Sweden}

\begin{abstract}
We show that the modulation of the phases of the laser beams of ultra-short pulses leads to modulation of the two photon fluorescence intensity. The phase modulation technique when used in multi-photon microscopy  can improve the signal to noise ratio. The technique can also be used in multiplexing the signals in the frequency domain in multi-focal raster scanning microscopy. As the technique avoids the use of array detectors as well as elaborate spatiotemporal multiplexing schemes it provides a convenient means to multi-focal scanning in axial direction. We show examples of such uses. Similar methodology can be used in other non-linear scanning microscopies, such as second or third harmonic generation microscopy.
\end{abstract}


\maketitle 

Interaction of matter with coherent light field leads to excitation with unique phase signature. Different spectroscopic techniques have been developed that rely on the interferometry of the wave-packets in the material systems generated by a sequence of laser pulses with controlled relative phases~\cite{RICE_1991, WARREN_2003,MARCUS2006, MARCUS2007, KARKI_2014_NC}.
 In these experiments the systems under investigation are resonantly excited and the phases of the laser pulses are manipulated to select out certain pathways in the excitations that are of interest~\cite{WARREN_2003, MARCUS2007, KARKI_2014_NC}.
 Similar approach can be taken in non-resonant excitations, such as multi-photon absorption. We use phase modulated laser pulses to induce two photon absorption in CdSe quantum dots. We show that the intensity of the subsequent fluorescence from the quantum dots is modulated at characteristic frequencies. This allows us to separate the fluorescence arising from the different interaction pathways. We apply the technique in multi-photon microscopy (MM) to address the specific challenges related to simultaneous raster scanning of multiple points~\cite{ANDERSEN_2001, DENK_2005, SO_2007, SQUIER_2007, SQUIER_2013}.

\begin{figure}[htbp]
\centerline{\includegraphics[width=.8\columnwidth]{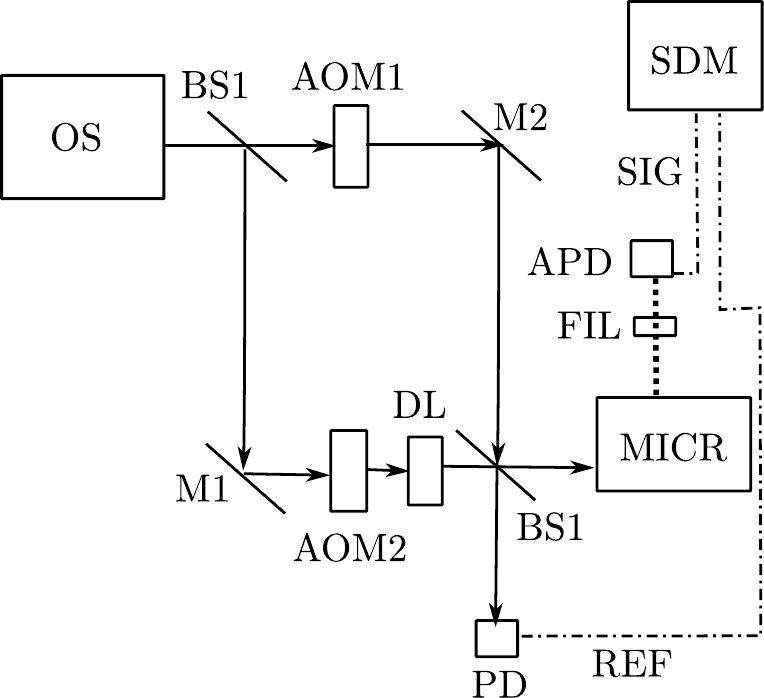}}
\caption{Schematic of the experimental setup. In the figure OS: Oscillator, BS: Beam splitter, M: Mirror, AOM: Acousto-optic modulator, DL: Delay line, PD: Photodiode, MICR: Microscope, FIL: Band pass filter (550$\pm$ 25 nm), APD: Avalanche ohotodiode,  SIG: Signal, REF: Reference and SDM: Signal demodulator. }
\label{FIG1}

\end{figure}

When using phase modulation in MM we split a pulsed laser beam from a femtosecond oscillator (Synergy from Femtolasers) into two replicas using a 50/50 beam splitter (Figure \ref{FIG1}). The frequencies of the pulses in the two beams are sheared by few tens of MHz by using acousto-optic modulators (AOMs). The electric fields $E_1(t)$ and $E_2(t)$ in a pair of pulses in the two beams can be written as $E_1(t) = A(t) \cos((\omega +\phi_1)t)$ and $E_2(t) = A(t-\tau) \cos((\omega+\phi_2)(t-\tau))$, respectively, where $A(t)$ is the envelope of the electric field, $\omega$ is the carrier frequency, $\tau$ is the time delay between the pulses, and $\phi_1$ and $\phi_2$ are the frequency shifts imparted by the modulation of the phase. The time delay between the pulses in a pulse pair can be adjusted by a delay line. For the best signal to noise contrast in the microscopy, we set $\tau$ close to zero. The electric field after another beam splitter that combines the two pulses collinearly is given by 

\begin{equation}\label{EQ1}
E_{\textrm{tot}}(t) = \frac{1}{2} A(t)\{\cos((\omega+\phi_1)t)+\cos((\omega+\phi_2)t)\}.
\end{equation}   

An $n$-photon absorption is 
 proportional to $(E_{\textrm{tot}}(t))^{2n}$ (the complete expression for $n$-photon absorption is given in the Appendix). The intensity of the resulting fluorescence gets modulated at the frequencies given by $f_{\textrm{mod}}= x \phi_1 \pm y \phi_2$, where $x$ and $y$ are  non-negative integers less or equal to $n$, i.e. $x\in \{0,1,2,...,n\}$ and $y \in \{0,1,2,...,n\}$. We set $\phi_1=55$ MHz and $\phi_2=55.1$ MHz in our experiments. The avalanche photodiode (APD) used in the setup has a bandwidth of 2 MHz. Consequently only the fluorescence modulated at the multiples of the relative frequency ($\Delta\phi=\phi_2-\phi_1$) can be resolved in our measurements. The amplitude as well as the phase of the signal can be obtained using a signal demodulator such as a lock-in amplifier. We use generalized lock-in amplifiers for this purpose\cite{KARKI2013A,KARKI2013C,KARKI2014_REV_SCI}. 

\begin{figure}[htbp]
\centering
\includegraphics[width=9cm]{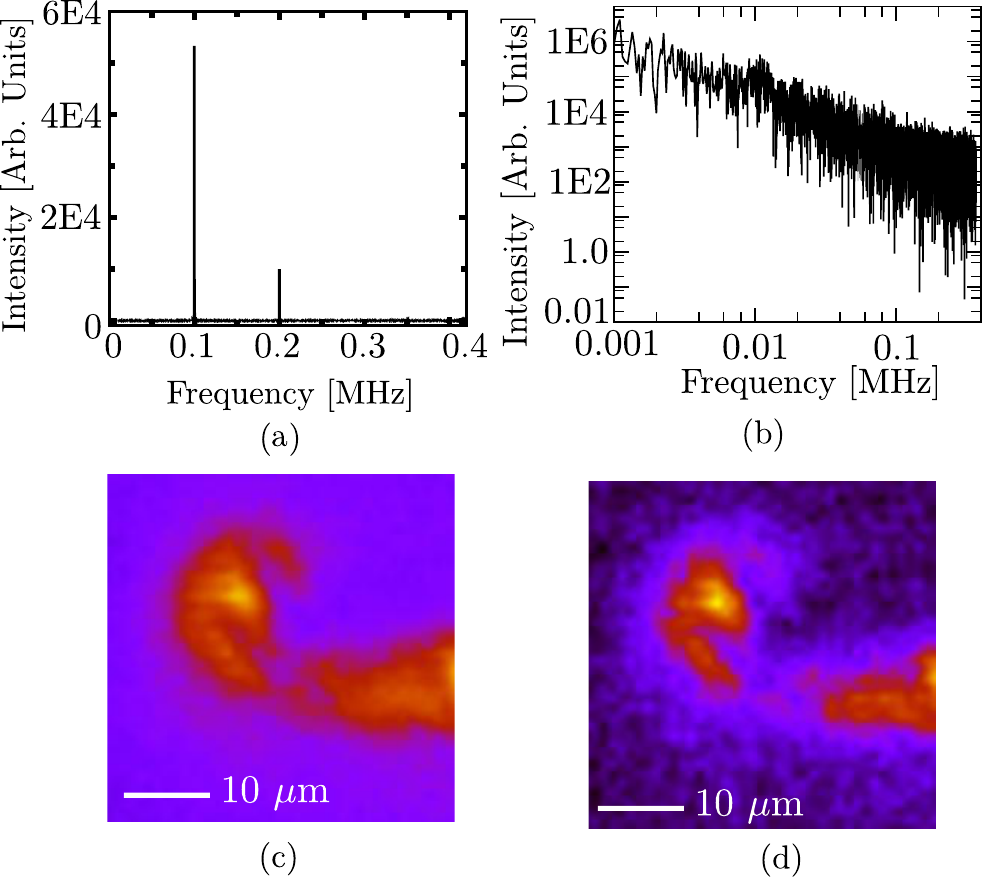}
\caption{(a) Fourier transform of the two photon fluorescence from CdSe quantum dots recorded for 10 ms. The phases of the two beams used to excite the fluorescence are modulated at 55 and 55.1 MHz, respectively. The resulting fluorescence  is modulated at 0.1 and 0.2 MHz. (b) noise of the detector at different frequencies. (c) raw image of a cluster of CdSe QDs taken by recording the normal two photon fluorescence image and (d) image of the same cluster taken by using the phase modulation of the beams.}
\label{FIG2}
\vspace{-0.5cm}
\end{figure}

The modulated intensity of the two photon fluorescence recorded in the measurements is given by 
\begin{equation}\label{EQ2}
I_{\textrm{det}}(t)\propto (4 \cos((\Delta\phi)t)+\cos(2(\Delta\phi)t)).
\end{equation} The fluorescence is separated from the excitation beams by using a band pass filter with pass band at 550$\pm$ 25 nm. The center wavelength of the excitation beams is at 780 nm and the full width at half maximum of the spectrum is about 130 nm.
Figure \ref{FIG2}(a) shows the Fourier transform of the fluorescence of CdSe quantum dots recorded for 10 ms. The transform shows two peaks at 0.1 and 0.2 MHz, which correspond to the contributions at $\Delta\phi$ and 2$\Delta\phi$ frequencies, respectively. 
The two contributions arise from the two different field matter interaction pathways. Out of the four field interactions when two of the fields come from each of the beams, the fluorescence gets modulated at 2$\Delta\phi$, when three of the fields come from one of the beams and the fourth from the other beam the fluorescence gets modulated at $\Delta\phi$ (see the Appendix for detailed discussion). Equation \ref{EQ2} predicts that the contribution at $\Delta\phi$ is four times the contribution at 2$\Delta\phi$. The amplitude of the fluorescence at the two frequencies in the measurements (Figure \ref{FIG2}(a)) deviate slightly from this relation. Equation \ref{EQ2} is valid when the intensity of both the beams at the focus spot are equal. However, this is not always the case. In general the signal at $\Delta\phi$ is more than or equal to four times the signal at 2$\Delta\phi$ (see the Appendix for further discussion). 
The signals at both the frequencies, $\Delta\phi$ and $2\Delta\phi$, can be used in microscopy. Figure \ref{FIG2} (c) shows the raw data of the normal two photon fluorescence image of a cluster of CdSe quantum dots on a glass slide. The image is taken by raster scanning the focus spot over the sample. The dwell time per scan point is 10 $\mu$s.  We use a reflective objective (Adjustable ReflX Objective 36X/0.5NA IR, Finite, Edmund optics) mounted on an inverted microscope (Nikon Ti-S) to focus the laser beams onto the sample. The focal length of the objective is about 5 mm.  The fluorescence is collected in the epi-direction. A multiphoton short-pass dichroic beamsplitter with the edge at 670 nm (FF670-SDi01-25x36, Semrock) separates the excitation beam from the collected fluorescence. The band-pass filter at 550 nm (OD 4) before the detector further reduces the chance of excitation photons reaching the detector.  Figure \ref{FIG2} (d)  shows the raw image of the same cluster using phase modulation. The image is recorded at the modulation frequency of 0.1 MHz. Comparing the two images, we see that the image taken with phase modulation has reduced noise.
The reduction in the noise is mainly due to the fact that at low frequencies the signal is heavily contaminated by the so called ``pink'' noise ($1/f$ noise) that is prevalent in all the electronic systems. Figure \ref{FIG2} (b) shows the electronic noise of the signal detection system that include the APD, the cables and the digitizer  at different frequencies. 

Apart from the improvement in the signal to noise ratio, a major advantage of using phase modulation in MM is in multiplexing the raster scanning by collecting fluorescence from multiple excitation spots in the sample. Various techniques have been developed for multi-focal multiphoton microscopy (MMM) over the last decade~\cite{HELL_1998, ANDERSEN_2001,SO_2007, SQUIER_2007}. The different techniques have their strengths, such as high speed imaging~\cite{HELL_1998}, and weaknesses, such as inhibited depth scanning~\cite{SO_2007}. The most recent advances include temporal multiplexing of the signal~\cite{SO_2007} to reduce the effect of interferences between the spatially overlapping foci, and remote focusing~\cite{WILSON_2008, SQUIER_2011} to allow rapid axial scanning. When applying phase modulation in MMM, we use more than one pair of beams to illuminate different spots on the sample. Each pair of the beams have different relative modulation frequencies, $\Delta\phi_1$, $\Delta\phi_2$, .... The fluorescence from all the spots are directed to a single element detector such as a photo-multiplier tube or an APD. The use of single element detector eliminates the limitations on the axial resolution which is inherent in imaging techniques that use array detectors~\cite{SO_2007,SQUIER_2009}. The signals from the different spots can be isolated by demodulating them at their respective modulation frequencies. This method of multiplexing signal in the frequency domain has advantages over the methods that use multiplexing in the time domain. Usually for the temporal multiplexing, the time between the different beams need to be delayed by the response time of the detector, which in most of the cases is a couple of nano-seconds~\cite{SO_2007}. This requires a rather elaborate optical setup or a specialized light source. Moreover, only few beams can be multiplexed within a repetition period of about 10-12 ns of an oscillator.

\begin{figure}[htbp]
\centering
\includegraphics[width=8cm]{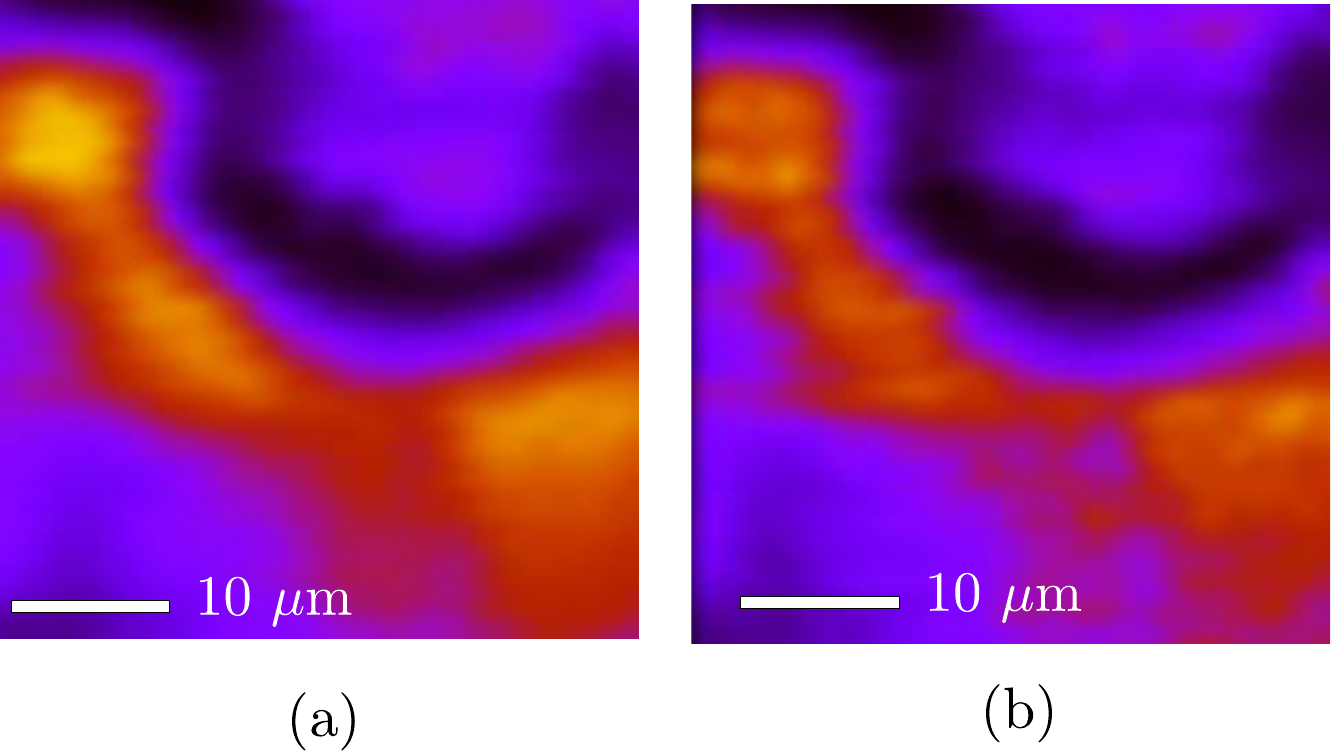}
\caption{Multi-point raster scanned images of a cluster of CdSe quantum dots on a glass slide. Image (b) is acquired at a depth of 4 $\mu$m below image (a). The relative phase modulations are set to $0.1$ and $0.17$ MHz, respectively. The images themselves are acquired at twice the relative modulation frequencies. }
\label{FIG3}
\vspace{-0.3cm}
\end{figure}

Figure \ref{FIG3} shows two images of a cluster of CdSe quantum dots. The two images are produced by raster scanning two pairs of beams. The relative phase modulations are set to $\Delta\phi_1=0.1$ and $\Delta\phi_2=0.17$ MHz, respectively. Only the images acquired at 2$\Delta\phi_1=0.2$ (top) and 2$\Delta\phi_2=0.34$ MHz (bottom) are shown. The two pairs of the beams are focused at different axial position (depth) separated by 4 $\mu$m on the sample. The distance between the focal points is controlled by remote focusing (i.e. by changing the divergence of one of the pairs of the beams with respect to the other). 

\begin{figure}[htbp]
\centering
\includegraphics[width=8cm]{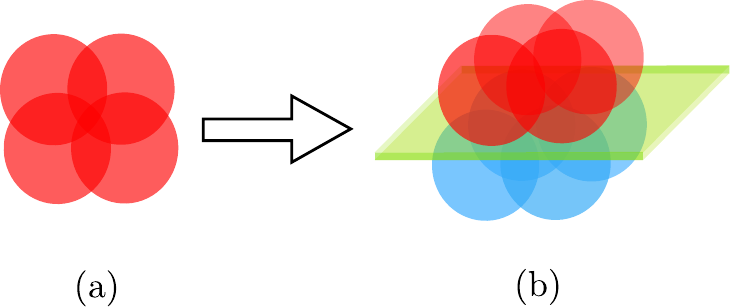}
\caption{A two dimensional grid (a) and a three dimensional grid (b) of beam patterns with over-lapping regions that can be used in multi-point raster scanning. (a) has four overlapping regions and (b) has 12 overlapping regions. The number of overlapping regions increases with the dimensionality of the beam pattern. }
\label{FIG4}
\vspace{-0.5cm}
\end{figure}

There is a technical difference between the normal and the phase modulated MMM in the way information is acquired from each raster scanned point. In a normal MMM one avoids any spatial and temporal overlap of the different beams~\cite{SQUIER_2009}. In the phase modulated MM one gets the information from the overlapping regions. This difference could be exploited to increase the number of points from which the signal is sampled. Instead of focusing two pairs of beams (four beams altogether) to two separate foci one can also use a 2D beam pattern as shown in Figure \ref{FIG4}(a) where the signal from four overlapping regions can be recorded at once. In general a square grid of $m\times n$ beams can have $2mn-m-n$ overlapping regions. The number of over-lapping regions can be increased further by using a 3D grid. Figure \ref{FIG4}(b) shows a $2\times 2\times 2$ grid of beams with 12 overlapping regions. In general a cubic grid of $m\times n\times o$ beams can have $3mno-mn-no-om$ overlapping regions that can be used in multi-point scanning in phase modulated MMM.

To conclude, we have shown how the phase modulation of the laser beams of ultra-short pulses can be used to modulate the intensity of the non-linear interaction with the sample. We have also shown that phase modulation can be used in MM to improve the signal to noise ratio and to multiplex the signals in the frequency domain for MMM. Though we have demonstrated the principle using two photon fluorescence, the phase modulation can be applied to any non-linear signal, such as second harmonic generation, to achieve similar advantages.

\textbf{Acknowledgements.}~~
Financial support from the Knut and Alice Wallenberg Foundation (KAW), the Swedish Research Council (VR), nanometer consortium lund (nmC@LU) and lund laser center (LLC) is gratefully acknowledged. 

\vspace{0.5cm}

\textbf{Appendix: Expression for multi-photon absorption using two phase modulated beams.} Let $\left( \omega + \phi_1 \right) t = a$
and $\left( \omega + \phi_2 \right) t = b$.  The expression for $n$-photon absorption is given by:
\begin{eqnarray*}
  E^{2 n}_{\textrm{tot}} \left( t \right) & = & \mathcal{E}^{2 n} \left( t
  \right) \left( 2 \cos \left( \frac{a + b}{2} \right)^{2 n} \cos \left(
  \frac{a - b}{2} \right)^{2 n} \right)\\
  & = & 2^{2 n} \mathcal{E}^{2 n} \left( t \right) \cos^{2 n} \left( x
  \right) \cos^{2 n} \left( y \right)
\end{eqnarray*}
with $x = \left( a + b \right) / 2, y = \left( a - b \right) / 2$. Using the trigonometric identities for the $n$-th power of cosine functions, we get:
\begin{small}
\begin{eqnarray*}
  E^{2 n}_{\textrm{tot}} \left( t \right) & \propto &  \left( C_n^{2 n} + 2 \sum_{k =
  0}^{n - 1} C_k^{2 n} \cos \left( \left( 2 n - 2 k \right) x \right) \right)\\
&&
  \left( C_n^{2 n} + 2 \sum_{l = 0}^{n - 1}
  C_l^{2 n} \cos \left( \left( 2 n - 2 l \right) y \right) \right)\\
  & = &  \left( C_n^{2 n}
  \right)^2 +  2 C_n^{2 n} \sum_{k =
  0}^{n - 1} C_k^{2 n} \cos \left( \left( 2 n - 2 k \right) x \right) +\\
   &&2 C_n^{2 n} \sum_{k = 0}^{n - 1} C_l^{2
  n} \cos \left( \left( 2 n - 2 l \right) y \right)\\
&& + \frac{4}{2^{2 n}}
  \sum_{k = 0}^{n - 1} \sum_{l = 0}^{n - 1} C_k^{2 n} C_l^{2 n} \cos \left(
  \left( 2 n - 2 k \right) x \right) \cos \left( \left( 2 n - 2 l \right) y
  \right) .
\end{eqnarray*}
\end{small}
Only the fluorescence arising from the third term (term that has $(2n-2l)y$ as the argument of the only cosine) in the expression given above contains modulations that can be resolved by a slow detector. Reverting back to the original notations, we get
\begin{eqnarray*}
  I_{\det} \left( t \right) & \propto & \sum_{l = 0}^{n - 1} C_l^{2 n} \cos
  \left( \left( n - l \right) \left( \phi_2 - \phi_1 \right) t \right) .
\end{eqnarray*}
The expression simplifies to
\begin{eqnarray*}
  I_{\det} \left( t \right) & \propto & \cos \left( 2 \left( \phi_2 - \phi_1
  \right) t \right) + 4 \cos \left( \left( \phi_2 - \phi_1 \right) t \right) 
\end{eqnarray*}
for the fluorescence due to two photon absorption.

When the intensities of the two beams are not the same we can follow the
similar derivation with different envelopes of the fields, $A_1\left( t \right)$ and $A_2 \left( t \right)$. The expression for the
resulting fluorescence intensity is given by
\begin{eqnarray*}
  I_{\det} \left( t \right) & \propto & A_1^2 \left( t \right)
  A_2^2 \left( t \right) \cos \left( 2 \left( \phi_2 - \phi_1
  \right) t \right)  +\\
&& 2 A_1 \left( t \right) A_2^3
  \left( t \right) \cos \left( \left( \phi_2 - \phi_1 \right) t \right) +\\
&&
  2A_1^3 \left( t \right) A_2 \left( t \right) \cos \left(
  \left( \phi_2 - \phi_1 \right) t \right),
\end{eqnarray*}
which shows that when three of the fields interacting with sample are from one
of the beams and the fourth field is from the other beam, the two photon
fluorescence gets modulated at $\phi_2 - \phi_1$. Similarly, when two of the
fields from each of the beams participate in the non-linear absorption, the
fluorescence gets modulated at 2$\left( \phi_2 - \phi_1 \right)$. 

If we take the general identity 
\begin{equation*}
(A_1(t)-A_2(t))^2\geq 0
\end{equation*}
then we get 
\begin{equation*}
A_1^2(t)+A_2^2(t) \geq 2 A_1(t) A_2(t).
\end{equation*}
As $A_1(t)$ and $A_2(t)$ are non-negative real numbers, multiplying both the sides by $2 A_1(t) A_2(t)$ gives
\begin{equation*}
2A_1^3(t)A_2(t)+2A_1(t)A_2^3(t)\geq 4 A_1^2(t) A_2^2(t).
\end{equation*}
 So in general, the fluorescence modulated at $\phi_2-\phi_1$ is more or equal to four times the fluorescence modulated at 2$(\phi_2-\phi_1)$.

\end{document}